\documentstyle[12pt,epsbox]{article}
\author{Ken Sekimoto\\
Yukawa Institute for Theoretical Physics, Kyoto University, 
Kyoto 606-01, Japan}
\title{Energetics of thermal ratchet models}

\begin{document}

\maketitle
\begin{abstract}
Several stochastic models called {\it thermal ratchet models} 
have been recently proposed to analyze 
the motor proteins such as myosin and kinesin,etc.
We propose a method to study the energetics of those models, 
and show how 
the rate of energy consumption and the energy dissipation are evaluated.
As a demonstration we consider "Feynman's ratchet", a typical 
fluctuating heat engine.
\end{abstract}



The motor proteins, such as myosin or kinesin, act as the energy transducer
in our lives, i.e., they produce mechanical work as they consume
chemical energy.
Motivated by such systems various phenomenological models have recently 
been proposed.~\cite{VO,Pr,Pe,Ma,As}
These models, which are called {\it thermal ratchet models}, share the common 
feature that the net work is obtained by rectifying the random thermal 
fluctuation of the system and that the energy of chemical reaction is 
implicitly supposed to be consumed for operating the rectifier, but not for
driving the system directly.

By now there are no systematic study of the energetics of those models. 
In most of the literatures the model stochastic equation (see 
eq. (\ref{Langevin}) below) or the equivalent Fokker-Planck equation~\cite{Ga} 
have been solved, and the average power output has been calculated, while
no assessment has been done to the total consumption of the energy and/or 
the energy wasted into heat bath. 
To the author's knowledge the only literatures on thermal ratchet model
which refers the energetics are the following:
Feynman~\cite{Fe} invented what is called Feynman's 
ratchet today and and analyzed in his textbook the energetics in a 
qualitative argument.
Magnasco~\cite{Ma2} considered so-called Szilard's heat engine and claimed
an expression of the net power consumed by the machine 
in steady state (see eq.(\ref{avD}) below).
Also there is a proposal of the formula of the total energy 
consumption~\cite{Pr} (with no derivation) which is variant from our result.

It is of much interest to study the energetics of biological systems 
since, in actuality, some of the motor proteins is reported to have very high 
efficiency of energy conversion.~\cite{Ya}
Form theoretical point of view, the formalization of energetics
is motivated in relation to the question: ``what should
the comprehensive phenomenological model of motor protein be?''
If a model of motor protein incorporates explicitly the chemical
reaction processes, that is, binding of adenosine triphosphate (ATP), 
hydrolysis of the ATP and releasing of the hydrolysis products, then
the energetics  such as the efficiency of energy conversion should be
tested within the model, and the framework given below will play a 
decisive role for such analysis.
The existing models of thermal ratchets, however, do not yet fully 
incorporate the reaction dynamics, but merely assume that the reaction 
dynamics is somehow correlated to the dynamics of the model.
The analysis of the energetics of such model and its comparison with 
real experimental results should, therefore, serve to judge
how substantial part of the energetics is grasped within the model.

In regard to the present status mentioned above, we would like to present 
an approach to the energetics of stochastic models of thermal ratchet
or, more specifically, to show how one can define the rate of total energy 
consumption or the rate of energy dissipation into heat.
The main purpose of this Letter is to describe a basic framework
and we will not exhaust the application to all the existing models. 
After establishing the framework the latter task can be
done in principle with the knowledge of the pertinent probability 
distribution functions which are already given in the literatures.
Below we firstly describe the general idea, and then 
we specify it for three typical categories of the existing models. 
After that we take up a version of so-called Feynman's
ratchet model,~\cite{Fe} as a demonstration, 
and show several concrete results of energetics.

{\it General argument:}\
In the thermal ratchet models, the whole system consists of the following
four parts:\\
 (i) the energy transducer, whose state variable is denoted by $x$, which may
be generalized to be more than one degree of freedom,\\
(ii) the external system, whose state variable is denoted by $y$, \\
(iii) the heat bath, and \\
(iv) the load $L$ to which the transducer does work.\\
The interaction of the transducer with the external system and the
one with the load are assumed to be potential-like with the potential 
$U(x,y)\equiv U_0(x,y)+L x$, where $U_0$ is a periodic function of $x$
with a period $\ell$, i.e., $U_0(x+\ell,y)=U_0(x,y)$.
The interaction with the heat bath is treated stochastically so that 
the heat bath
exerts an instantaneous force  $-\gamma {dx}/{dt}+\xi(t)$, where
$\gamma$ is the friction constant and $\xi(t)$ is, as usual, the white Gaussian
process characterized by the ensemble averages, $\langle\xi(t)\rangle=0$ and 
$\langle\xi(t)\xi(t')\rangle=({2 \gamma}/{\beta})\delta(t-t')$ with 
$\beta={(k_{\rm B}T)}^{-1}$ being the inverse of the temperature of the bath
times Boltzmann constant.
The external system is assumed to be statistically independent of $\xi(t)$.

The dynamics of the system is described by the following Langevin equation;
\begin{equation}
-\frac{\partial U(x,y)}{\partial x}+\left[ 
-\gamma \frac{dx}{dt}+\xi(t)\right]=0, \quad x(t_i)=x_i.
\label{Langevin}
\end{equation}
 The property of the external system is not specified at this point 
except that the variation of $y$ leads to a bounded variation for $U$, 
that is, whether $y(t)$ itself varies in a bounded region~\cite{Pr,As,Fe,Pr2}
or $U$ is assumed to be periodic with respect to $y$~\cite{Ma3}.

We introduce the three quantities concerning the energetics of the system:
{\it (a)} the work $W$ done by the transducer to the load during the period, 
say, $t_i<t<t_i$, which is formally given as
\begin{equation}
W=U(x(t_f),y(t_f))- U(x(t_i),y(t_i)),
\label{work}
\end{equation}
{\it (b)} the dissipation of energy $D$ to the heat bath, 
and {\it (c)} the total consumption of energy $R$ coming from the external 
system in the meantime. 
The low of energy conservation requires the relation $W+D=R$.
As noted before, there are many calculation of the first quantity in the form 
of average power, $\langle dW/dt\rangle=$ $L \langle dx/dt\rangle$,
where  $\langle dx/dt\rangle$ can
be directly obtained from the simulation of the Langevin equation 
(\ref{Langevin}) or by solving the equivalent Fokker-Planck equation 
and integrating the probability current over 
a section perpendicular to the $x-$axis.

Our reasoning for obtaining the remaining two quantities is based
on the following observation: 
The equation (\ref{Langevin}) embodies {\it the balance of forces acting on 
the transducer} and, therefore, the transducer exerts the reaction force 
$-\left[ -\gamma {dx}/{dt}+\xi(t) \right]$ to the heat bath.
Then, the dissipation, which is the work done by the transducer onto the
heat bath, is given as the following Stieltjes integral
\begin{eqnarray}
D&=&-\int_{t=t_i}^{t=t_f} 
 \left[ -\gamma \frac{dx}{dt}+\xi(t) \right] 
dx(t) \nonumber \\
 &=& \int_{t=t_i}^{t=t_f}
\left[ -\frac{\partial U((x(t),y(t))}{\partial x}\right] dx(t).
\label{dissipation}
\end{eqnarray}
To move on to the second line on the right hand side we have 
used (\ref{Langevin}). 
We should note here that the probability theory tells that the above
integrals should be interpreted as the stochastic one in the Storatonovich 
sense,~\cite{Ga} 
and that we can perform the usual integration rules such as integration
by parts or the change of integration variable. 
By the conservation law $W+D=R$ the total consumption of the energy becomes
\begin{equation}
R=\int_{t=t_i}^{t=t_f}
\left[ dU(x(t),y(t))-\frac{\partial U(x(t),y(t))}{\partial x} dx(t)\right]
\label{consumption}
\end{equation}
Below  we apply the above formula of $R$ 
to the three categories of models with different assumptions on the
variable $y(t)$, and show how this formula can be transformed into 
a physically appealing expression.

{\it Category1:  $y(t)$ is a given periodic function, 
$y(t+T)=y(t)$}.~\cite{Pr,Ma}
In this case we may write $U(x,y(t))$ as ${\cal U}(x,t)$ and
using the identity $d\, {\cal U}=$
$({\partial {\cal U}}/{\partial x}) dx+$
$({\partial {\cal U}}/{\partial t}) dt$, 
(\ref{consumption}) becomes,
\begin{equation}
R=\int_{t_i}^{t_f}
\frac{\partial {\cal U}(x(t),t)}{\partial t} dt, \quad (\mbox{\it Category 1}).
\label{case1}
\end{equation}
This expression tells that the input of energy from the external system is
done by lifting the potential $\cal U$ while the state of the 
transducer, $x$, is virtually fixed.
Suppose that the probability distribution function $P(x,t)$ 
is available as the solution of the Fokker-Planck equation which corresponds to
(\ref{Langevin}),  
${\partial P}/{\partial t}=-{\partial J}/{\partial x}$, 
where $J\equiv-\gamma^{-1}$
$(\beta^{-1}\partial P/\partial x+$ $P\partial U/\partial x)$
is the probability current, and the initial condition $P(x,0)=\delta(x-x_i)$
is satisfied. 
The average $\langle R\rangle$ is then given as 
\begin{equation}
\langle R\rangle = \int_{t_i}^{t_f} dt
\int_\Omega \frac{\partial {\cal U}(x,t)}{\partial t} P(x,t)dx,
\label{avR}
\end{equation}
where $\Omega$ is the domain of the variable $x$. 
Although we treat $x$ as a single degree of freedom 
for the simplicity of explanation,
the generalization to more than two degrees of freedom~\cite{DV} is
straightforward.
We can show after integration by parts that  (\ref{avR}) is equivalent 
to the following expression for $\langle D\rangle$, 
\begin{equation}
\langle D\rangle =\int_{t_i}^{t_f} dt
\int_\Omega dx\left[-\frac{\partial {\cal U}}{\partial x}\right] J,
\label{avD}
\end{equation}
Magnasco~\cite{Ma2} has discussed a similar expression 
in the case of steady state probability distribution,
for a time independent potential.

{\it Category2: $y(t)$ obeys discrete Markov process}.~\cite{Pr,Ma,As,DV}
The value of $x(t)$ is assumed to be continuous
upon the jump of $y(t)$ from one discrete value to the other.
Let us denote by $\{t_j\}$ the times at which such jumps occur 
upon a particular realization of $y(t)$.
Then  $R$ in (\ref{consumption}) becomes
\begin{eqnarray}
R = \sum_{j}[U(x(t_j),y(t_j+0))-&& U(x(t_j), y(t_j-0))], \nonumber \\  
&&(\mbox{\it Category 2}),
\label{case2}
\end{eqnarray}
where the sum is taken for all the jumps occurred during $t_i$ and $t_f$.
This result shows that, as in (\ref{case1}), the external system 
puts energy into the transducer by lifting the potential
with $x$ taking its instantaneous value.

In order to calculate the average $\langle R\rangle$, we introduce the 
transition probability of $y$: If we distinguish by $\{ \sigma \}$ the
possible discrete values of $y$, the probability distribution function,
of $x$ and $y$, 
$P_\sigma(x,t)$, obeys the Fokker-Planck equation like the following 
form,~\cite{Ma,As,Pr3} 
\begin{equation}
 \frac{\partial P_\sigma}{\partial t}= 
 -\frac{\partial J_\sigma}{\partial x}+
 \sum_{\sigma'}P_{\sigma'}W_{\sigma'\sigma}-
 \sum_{\sigma'}P_{\sigma}W_{\sigma\sigma'}, 
\end{equation}
where $J_\sigma \equiv-\gamma^{-1}$
$(\beta^{-1}\partial P_\sigma/\partial x+$ 
$P_\sigma \partial U_\sigma/\partial x)$ 
and $U_\sigma(x)$ denote, respectively, the 
probability current in the domain of $x$-variable, $\Omega$,
and the potential with $y$ taking its $\sigma$-th value, and
$W_{\sigma\sigma'}$ is the 
transition rate of $y$ from $\sigma$-th to $\sigma'$-th value. 
($W_{\sigma\sigma'}$ can be a function of $x$ and $t$.) 
Using $W_{\sigma\sigma'}$ and
the probability $P_\sigma(x,t)$ the average consumption $\langle R\rangle$
is given as,
\begin{eqnarray}
\langle R\rangle =\frac{1}{2}\int_{t_i}^{t_f}dt
\int_{\Omega}dx \sum_{\sigma}&& \sum_{\sigma'}
(P_\sigma W_{\sigma \sigma'}-P_{\sigma'}W_{\sigma'\sigma})
\times \nonumber \\
& &\times [U_{\sigma'}-U_\sigma],
\label{avR2}
\end{eqnarray}
where the argument $x$ has been suppressed. 
From this expression it is clear that 
the net consumption of the energy comes out from the lifting of potential
due to the breaking of the detailed balance with respect to $y$-values.

{\it Category3: $y(t)$ is a stochastic process influenced by the second 
heat bath}.~\cite{VO,Fe} $y(t)$ 
is assumed to obey the following Langevin equation,
\begin{equation}
-\frac{\partial U(x,y)}{\partial y}+\left[ 
-\hat{\gamma} \frac{dy}{dt}+\hat{\xi}(t)\right]=0, \quad y(t_i)=y,
\label{Langeviny}
\end{equation}
where the second heat bath is characterized by $\hat{\gamma}$ and 
$\hat{\beta} (\neq \beta)$ through the conditions, 
$\langle \hat{\xi}(t)\rangle=0$ and 
$\langle \hat{\xi}(t)\hat{\xi}(t')\rangle=(2\hat{\gamma}/\hat{\beta})
\delta(t-t')$.
The above equation tells the balance of forces on the degree of 
freedom, $y$, just as (\ref{Langevin}) does on $x$.
By the same reasoning as we derived eq. (\ref{dissipation}), the consumption
of energy from the second heat bath, $R$, is given as
\begin{eqnarray}
R &= & -\int_{t=t_i}^{t=t_f}
\left[-\frac{\partial U(x(t),y(t))}{\partial y}\right]  dy(t), \nonumber \\
&\equiv &-\hat{D}  \qquad \qquad (\mbox{\it Category 3}).
\label{case3}
\end{eqnarray}
Using the identity $dU=(\partial U/\partial x)dx+$ 
$(\partial U/\partial y)dy$ we can verify that $D+\hat{D}+W=0$ holds.

In the present case the above stochastic integral $R$ should be evaluated 
directly, not {\it via} the form like (\ref{case1}) above.
The evaluation of 
$\langle R\rangle$ ($=-\langle \hat{D}\rangle$ 
or $\langle D\rangle$), therefore, requires some
care about the Storatonovich calculus.~\cite{Ga} 
Noting that the probability distribution of $dx(t)$ and $dy(t)$ obeys
the Fokker-Planck equation with the initial condition specified at the 
time $t$, and that $\partial U/\partial x$ or  $\partial U/\partial y$
should be  evaluated at the midpoint, $t+dt/2$, as a rule of 
Storatonovich calculus, we arrive at the expressions; 
\begin{equation}
\langle {D}\rangle=\int_{t_i}^{t_f} dt
\int_\Omega dx \int_{\hat{\Omega}} dy
\left[-\frac{\partial { U(x,y)}}{\partial x}\right]{J},
\label{Dcase3}
\end{equation} 
\begin{equation}
\langle \hat{D}\rangle=\int_{t_i}^{t_f} dt
\int_\Omega dx \int_{\hat{\Omega}} dy
\left[-\frac{\partial { U(x,y)}}{\partial y}\right]\hat{J},
\label{barDcase3}
\end{equation} 
where $\hat{\Omega}$ is the range of the variable, $y$, and 
$\hat{J}\equiv$ $-\hat{\gamma}^{-1}$
$(\hat{\beta}^{-1}\partial P/\partial y+$ $P\partial U/\partial y)$ 
is the probability current of $y$. 
In the absence of the load, $L=0$, the transducer described by 
(\ref{Langevin}) and (\ref{Langeviny}) acts as a passive heat conductor.
Especially, for harmonic coupling $U(x,y)=(k/2) {(x-y)}^2$ with $k>0$, 
we can directly calculate the energy conduction rate; 
$-\langle d\hat{D}/dt\rangle=\langle dD/dt\rangle=$  
$k({\hat{\beta}}^{-1}-{\beta}^{-1})/(\gamma+\hat{\gamma})$.

{\it Two remarks:} 
Firstly the derivation of the expressions  (\ref{avD}), (\ref{Dcase3}) 
and (\ref{barDcase3}) from the stochastic integrals
(\ref{dissipation}) or (\ref{case3}) can be done more shortly but
symbolically by
regarding $\mbox{\boldmath$v$}\equiv$ $-{\gamma}^{-1}$
$({\beta}^{-1}\partial /\partial x+$ $\partial U/\partial x)$ or 
$\hat{\mbox{\boldmath$v$}}\equiv$ $-\hat{\gamma}^{-1}$
$(\hat{\beta}^{-1}\partial /\partial y+$ $\partial U/\partial y)$   
as the velocity operators along $x$ and $y$ directions, respectively,
and rewriting, for example, $\int_{t=t_i}^{t=t_f} \langle
[-\partial U/\partial x]dx(t)\rangle $ as 
$\int_{t_i}^{t_f} dt$
$\langle [-\partial U/\partial x] \mbox{\boldmath$v$}\rangle$, etc.

Secondly  in the above description 
the probabilities and the currents have been assumed to be the solutions
of the initial value problem with a definite initial value of $x$ (and of 
$y$) at the initial time $t=t_i$.
We can show, however, that in the long-time limit  
the integrals in (\ref{avR}), (\ref{avD}),
(\ref{avR2}), (\ref{Dcase3}) and (\ref{barDcase3})  
may be alternatively evaluated by using the solutions of  
Fokker-Planck equation which are periodic along $x$-direction with
the period $\ell$
(and, for {\it Category 1}, along $t-$direction with the period 
$T$)
under the normalization condition imposed within the spatial period.  
If we use such solutions, the integral $\int_\Omega dx$ in the equations 
mentioned above should be replaced by that over the period $\ell$, say, 
$\int_0^{\ell} dx$.

{\it Example: Feynman's ratchet}.~\cite{Fe}
Feynman invented a thoughtful heat engine consisting of a {\it ratchet} wheel
joined tightly to a rotatable vane immersed in a first heat bath,
and a {\it pawl} that is loosely attached by an elastic spring 
to the ratchet's tooth and is immersed in a second heat bath.
The profile of the tooth of the ratchet is asymmetric and the temperatures
of the two heat baths are made different.
It has been shown~\cite{Fe} that in this non-equilibrium system 
the ratchet wheel can generate a torque even under a load.
Depending on the temperatures of the heat baths,  one of the baths 
acts as a source of energy, while 
the other bath acts as a breaking media, or a cooling media,
which absorbs the kinetic energy of the part immersed therein.
If the effect of inertia is neglected, his model is a typical example of
the transducer of the {\it Category 3} described above.
Since Feynman has left only qualitative discussion on his 
model, the concrete description given below would be of some 
interest to demonstrate the feasibility of our framework.

\begin{figure}[t]
\begin{minipage}{8cm}
   \postscriptbox{8cm}{8cm}{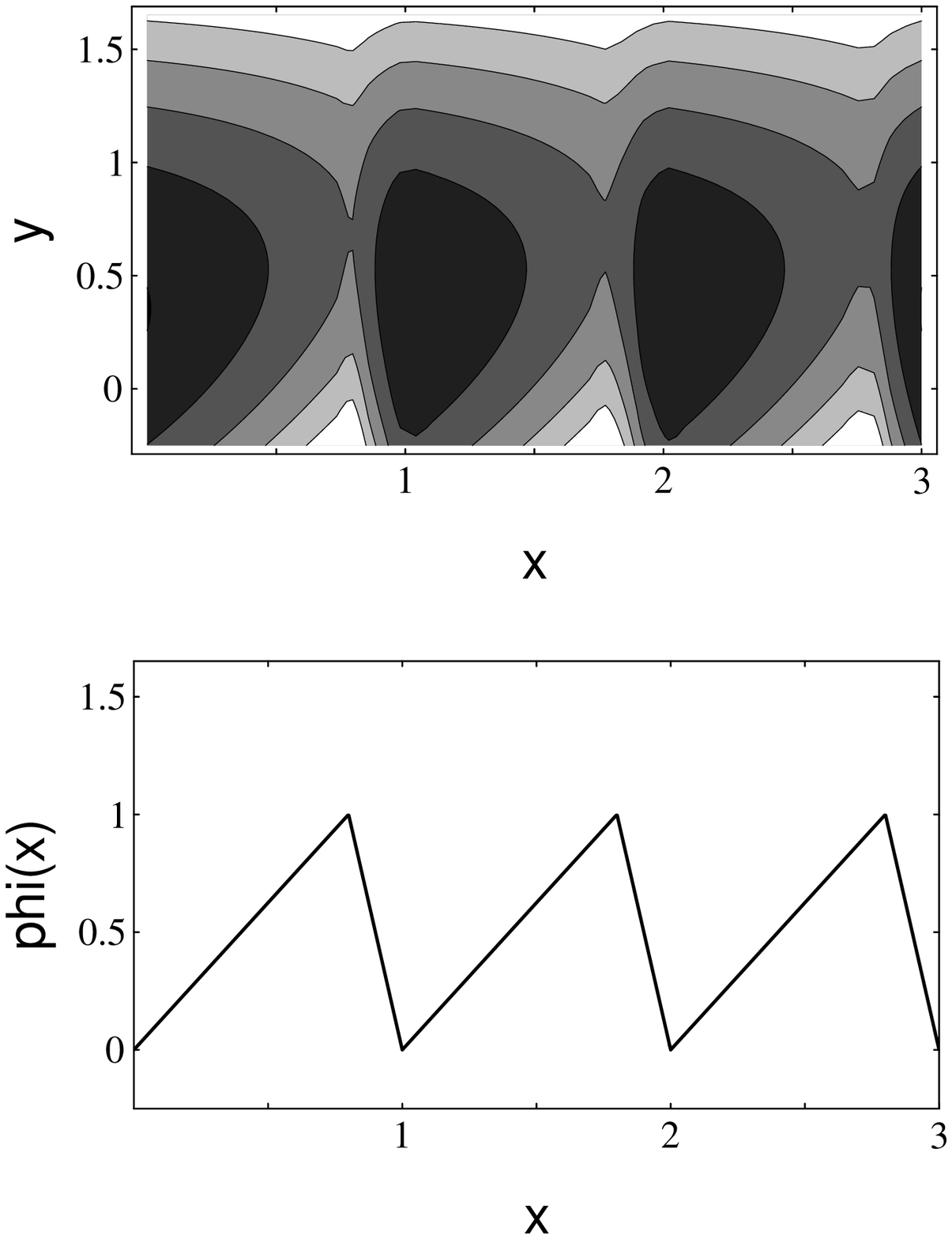}
   \caption{
In the {\it top} figure the contour plot of the potential $U(x,y)$ is shown.
The parameters we have chosen for (\protect\ref{ratchet}) in the text
are such that
$U_1(z)=e^{-z}$ and $U_2 (y)=\frac{1}{2}y^2$, and the profile of $\phi(x)$ is
given in the {\it bottom} figure.
The brighter region indicates the higher
potential, and the spacing between the contours corresponds to the height of
$0.5$. 
We have descretized the region of $0<x<\ell\equiv 1$ and 
$-0.25<y<1.65$ into
$50 \times 35$ points and imposed the periodic boundary condition
at $x=0$ and $x=\ell$. 
As for the boundaries $y=-0.25$ and $y=1.65$ we required 
$\hat{J}$ to vanish so that the probability is conserved within the 
region of calculation. 
The inverse temperatures and the friction constants are chosen to
be $\beta=2,$ $\hat{\beta}=4,$ $\gamma=1$ and $\hat{\gamma}=1$.
\label{fig:potential}}
\end{minipage}
\hspace{0.5cm}
\begin{minipage}{8cm}
   \postscriptbox{8cm}{8cm}{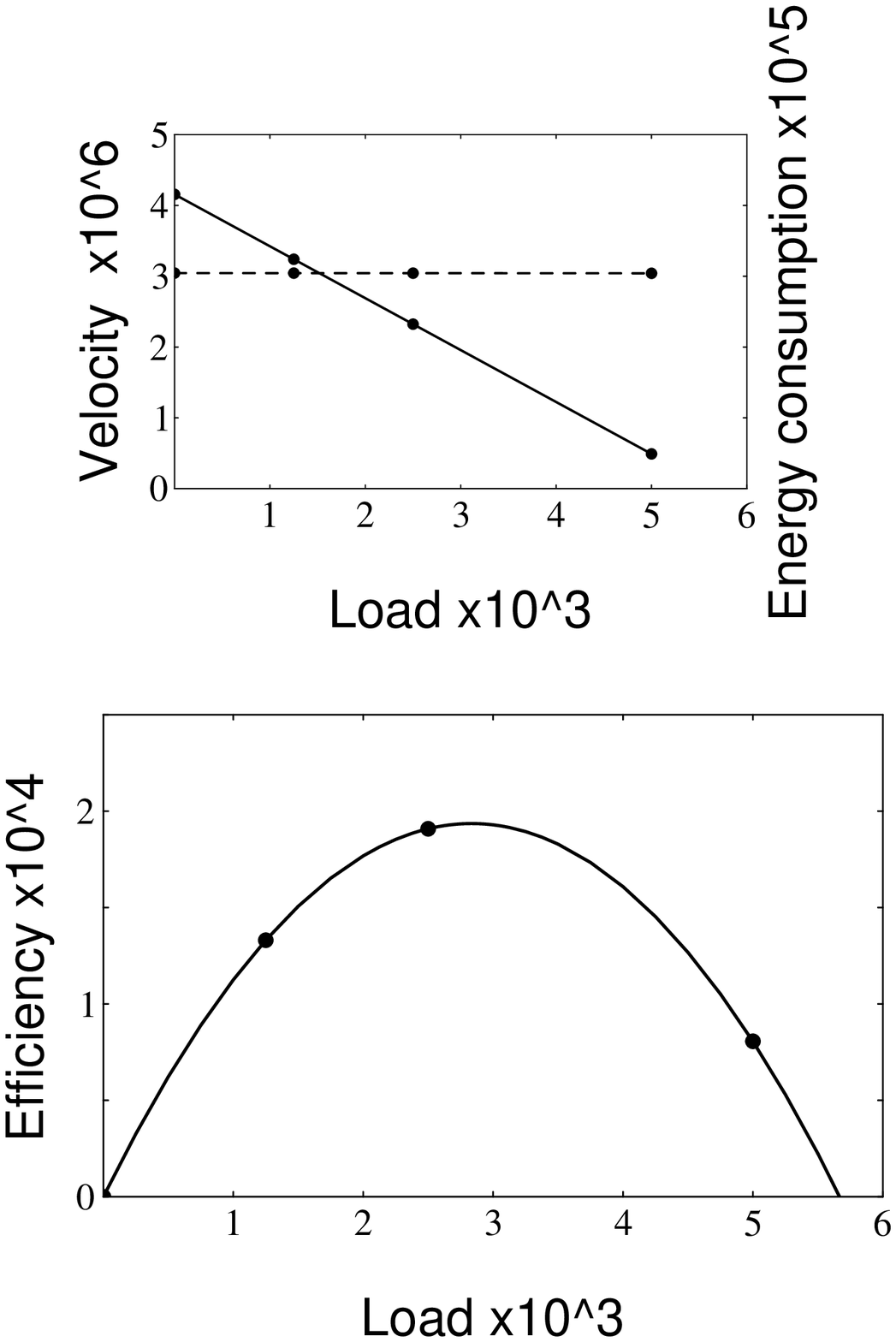}
   \caption{
{\it Top:} The mean velocity $\langle \frac{dx}{dt}\rangle$
(the dots tied by solid line, and the left ordinate) and 
the mean energy consumption rate $\langle \frac{dR}{dt}\rangle$
(the dots tied by dashed line, and the right ordinate)
as function of the load, $L$. The thick dots are the calculated points.
{\it Bottom:} Calculated efficiency $\eta$ (the thick dots)
and the extrapolated efficiency curve obtained from 
the quasi-linear dependence of both $\langle \frac{dx}{dt}\rangle$
and $\langle \frac{dR}{dt}\rangle$ on the load.
\label{fig:power}}
\end{minipage}
\end{figure}

The valuable $x$ in our notation
corresponds to the angle of rotation of a ratchet wheel
and $y$ represents the displacement of the pawl. 
The potential $U(x,y)$ can be given in the following form,
\begin{equation}
U(x,y)=U_{\rm 1}(y-\phi(x))+U_{\rm 2}(y)+Lx,
\label{ratchet}
\end{equation}
where $\phi(x)$ is the periodic function with a period $\ell$ that 
represents the asymmetric saw-tooth profile of the ratchet
and $U_{\rm 1}(z)$ stands for
the short-range repulsion between the pawl and the ratchet.
In the original model it is hard-core like; $U_{\rm 1}(z)=\infty$ 
for $z<0$ and $=0$ for $z\geq 0$. 
The second term $U_{\rm 2}(y)$ is the potential devised so that
the pawl is elastically pressed down onto the ratchet tooth.
We show in Fig.~\ref{fig:potential} the contour plot of the potential
energy $U(x,y)$ as well as the profile of $\phi(x)$,
which we used in our calculation. The values of the other parameters are also 
given in the figure caption.

As noted in the above {\it remarks}, in order to evaluate $\langle D\rangle$ or
$\langle \hat{D}\rangle$ we need only the stationary and periodic 
solution of the Fokker-Planck equation 
${\partial P}/{\partial t}=-{\partial J}/{\partial x}
-{\partial \hat{J}}/{\partial y}$, normalized within the range,
$0\leq x\leq \ell$. 
Using such solution, $\langle dD/dt\rangle$ is given as
\begin{equation}
\left\langle \frac{dD}{dt}\right\rangle =\int_0^\ell dx\int_{\hat{\Omega}} dy
\left[ -\frac{\partial U}{\partial x}\right] J,
\end{equation}
and $\langle d\hat{D}/dt\rangle$ is obtained similarly.
The average velocity $\langle dx/dt\rangle$ can be calculated as
$\ell \int_{\hat{\Omega}}dy J$. 
The stationarity condition 
$\partial J/\partial x+\partial \hat{J}/\partial y=0$
assures that the last integral is independent of the variable $x$.

The result of our numerical calculation is given in Fig.~\ref{fig:power}. 
In the top figure we show the mean velocity $\langle \frac{dx}{dt}\rangle$
and the mean energy consumption rate for four values of the load.
The efficiency $\eta$ can be calculated from these data as
$\eta\equiv L\langle \frac{dx}{dt}\rangle/\langle \frac{dR}{dt}\rangle$ and
is shown in the bottom figure of Fig.~\ref{fig:power}. 
Since both $\langle \frac{dx}{dt}\rangle$ and $\langle \frac{dR}{dt}\rangle$
depend almost linearly on the load, we can safely extrapolate to find that
the velocity and the efficiency vanishes when the load is about $0.0057$.
In the steady states $J$ and $\hat{J}$ generally compose a finite circulation
$\partial J/\partial y-\partial \hat{J}/\partial x$
and, as Magnasco has pointed out,~\cite{Ma} 
these states are qualitatively different from equilibrium states
even if $\left\langle \frac{dx}{dt}\right\rangle$ vanishes; 
the coupling between $x$ and $y$ allows the transfer of energy even without
doing work.
The efficiency of the model turned out to be very small. 
It is because  we have chosen a rather moderate 
potential variation  (the difference of $\beta U(x,y)$ between the minima 
and the saddle points is about one) 
in order to assure the numerical accuracy.
The feasibility of our formalism, however, should be understood by the 
present example.

In this Letter we have developed the framework to analyze the energetics of
thermal ratchets that work as energy transducers while keeping contact with 
heat bath(s).
We have developed here the point of view that
Langevin equations imply the balance of forces and that the energetics of 
the system can be analyzed based on this balance relation 
with the aid of a standard probability theory.
We have not exhausted the possible application of our scheme; for example,
the external system may be a chaotic dynamical system.~\cite{Ho}
It is a future task to construct the comprehensive phenomenological model 
of motor proteins in which we should specify a biochemically correct expression
of the external system.

\section*{Acknowledgements}
The author gratefully acknowledges T. Hondou and K. Sato for fruitful
discussion. This work has been supported partly by Asahi Glass Foundation.

\end{document}